\documentclass[12pt]{iopart}

\expandafter\let\csname equation*\endcsname\relax
\expandafter\let\csname endequation*\endcsname\relax
\usepackage{amsmath}
\UseRawInputEncoding
\usepackage{graphicx}
\usepackage{subfigure}
\usepackage[ruled, linesnumbered]{algorithm2e}
\usepackage{setspace}
\usepackage{array}
\usepackage{booktabs}
\usepackage{tabularx}
\usepackage{multirow}
\usepackage{hyperref}
\usepackage{cleveref}

\usepackage{bm}
\hypersetup{hypertex=true,
	colorlinks=true,
	linkcolor=blue,
	anchorcolor=blue,
	citecolor=blue}
\usepackage{graphbox}

\usepackage{appendix}

\begin{document}

\title[]{A scattering correction method for CT reconstruction based on the Wavelet Adaptive Material-dependent Boltzmann Transport Equation (WAM-BTE)}

\author{Huiying Pan$^{1,\star}$, Xu Jiang$^2$, and Xing Zhao$^{2,*}$}
\address{$^1$ School of Mathematics and Statistics, Nanjing University of Information Science and Technology, Nanjing, 210044, China}
\address{$^2$ School of Mathematical Sciences, Capital Normal University, Beijing, 100048, China}
\ead{$\star$ phy$\_$1995@126.com}
\ead{* zhaoxing$\_$1999@126.com}

\vspace{10pt}
\begin{indented}
\item[Received] xxxxxx
\end{indented}

\begin{abstract}
X-ray computed tomography (CT) is an essential imaging technology in clinical diagnosis. However, scattered photons can reduce image contrast and introduce CT value bias, which severely degrades image quality. Recently, scatter correction methods based on the Boltzmann transport equation (BTE) have attracted increasing attention due to their high physical accuracy and flexibility. Nevertheless, existing BTE-based methods usually employ single-material models, which cannot accurately describe the nonlinear energy dependence of photon interaction cross-sections in different materials. In this work, a scatter correction method based on the wavelet adaptive material-dependent BTE (WAM-BTE) is proposed. The conventional single-material model is extended to a multi-material model by introducing material-dependent scattering distributions. Furthermore, an adaptive multi-scale framework is established through wavelet decomposition. The low-frequency wavelet coefficients are used for coarse-scale scatter estimation to reduce computational complexity, while the high-frequency wavelet energy of high attenuation materials is utilized for adaptive local refinement. Theoretical analysis demonstrates that, when using the Haar basis function, the low-frequency wavelet coefficients at the $w$-th level are mathematically equivalent to block-average downsampling with a scale factor of $2^w$, except for a deterministic normalization factor. Experimental results show that the proposed WAM-BTE method achieves comparable accuracy to the Monte Carlo method while preserving the computational efficiency of coarse-scale estimation. The scatter calculation time for a single projection view is reduced to the millisecond level.

\vspace{1pc}
\end{abstract}

\section{Introduction}
\label{introduction}

X-ray computed tomography (CT) is an essential imaging technology in clinical diagnosis due to its non-invasive, intuitive, rapid, and cost-effective characteristics. However, X-ray photons inevitably interact with scanned objects through Compton scattering, Rayleigh scattering, and other physical processes. These interactions generate scattered photons, causing the measured projections to deviate from the conventional Beer-Lambert model based on primary photons. Consequently, scatter artifacts, including cupping artifacts, streak artifacts, and CT value bias, severely degrades image quality.

Various scatter correction methods have been developed, including hardware-based and software-based approaches. Hardware-based methods aim to reduce the influence of scattered photons by preventing them from reaching the detector, such as air gaps methods \cite{airgap}, anti-scatter grids methods \cite{anti-scatter}, beam blocker methods \cite{beamblocker}, and so on \cite{bowtiefilter,pre-grids}. However, these approaches usually require hardware changes, which increase system costs and reduce the flexibility of scanning configurations. Therefore, software-based correction methods are more commonly used in practical CT systems.

Software-based methods usually estimate the scatter distribution from measured data, and then subtract the estimated scatter signal from the measurements to obtain corrected data. Existing software-based methods can be roughly categorized into four groups: Monte Carlo (MC) methods, scatter-kernel-based (SK) methods, deep learning (DL) methods, and Boltzmann-transport-equation-based (BTE) methods.

MC methods \cite{MC1,MC2} simulate the complete physical process of X-ray photon transport, including interactions with the scanned objects, photon attenuation, and detector detection. The scatter distribution is obtained by statistical information on energy deposition and interaction events. Due to their strong physical interpretability and high theoretical accuracy, MC methods are considered the "gold standard" for scatter correction \cite{MCgold}. However, MC methods suffer from high computational costs and slow simulation speed \cite{DSEnet,MCAcc1,MCAcc2}. Although some acceleration strategies have been developed, they still do not limits their applications of real-time imaging. Moreover, practical CT systems often involve complex scanning conditions, and critical system parameters are difficult to obtain accurately. The uncertainty of these parameters affect the reliability of estimated scatter distribution.

SK methods approximate scatter signals by convolving the signal with multiple Gaussian kernels. The kernel functions are typically designed according to two properties derived from the ideal Klein-Nishina (K-N) formula, the "rotational symmetry" of azimuthal scattering distribution and the "radial attenuation property" with increasing incident angle. The SKS method \cite{sks} first introduced Gaussian kernels for scatter estimation, while PolySKS \cite{polysks,sjnOLEN} further extended this framework to multi-energy conditions. The fSKS or faSKS method \cite{fsks} takes into account the nonstationary kernels and improves model accuracy. These methods require fewer parameters and provide fast computation, making them widely used in practical systems. However, the kernel assumptions are highly simplified. The electron binding effects at low energies lead to deviations of Compton scattering from the K-N formula, and the scatter profile may even lose its central peak characteristics. Therefore, the precision of is limited.

DL methods learn the mapping relationship between scatter-contaminated and scatter-free images. After training with large-scale paired datasets, these methods can achieve effective scatter correction performance. Some methods directly learn the mapping function in image domain \cite{FSTUnet,Syntheticnet}. Other methods employ U-Net or generative adversarial network (GAN) to estimate scatter in projection or photon domain \cite{ScatterNet,cycleGAN,Pix2pixGAN}. Moreover, some methods embed traditional scatter models within deep neural networks to improve image quality \cite{deepsplines,PIDL,KAN}. However, scattered signals are highly dependent on object structures and material compositions. When the distribution of practical data differs significantly from training data, the generalization capability of deep learning methods may degrade substantially. Furthermore, due to their black-box characteristics, these methods provide limited physical interpretability and may introduce over-correction artifacts.

BTE methods describe the transport behavior of photons in phase space, allowing scatter signals to be modeled from a physical perspective \cite{BTEaccuracy}. Acuros CTS \cite{AcurosCT1} discretizes the phase space in multiple dimensions and estimates full-order scattering through voxel-based ray tracing and forced detection integration. The UBES method \cite{UBES} expanded the linear BTE into a hierarchical equation system using perturbation theory, and presented a semi-analytical fast algorithm. MSD-LBTE \cite{MSD-LBTE} further introduces the energy dimension into phase space and applies phase-space matrix factorization, enabling simultaneous estimation of scatter distributions for multiple spectra in a single BTE computation. The BTE methods have characteristics of high accuracy, good generality, and excellent physical interpretability, and received increasing attention recently. Some researches focuse on the discrete forms of phase space \cite{BTEdiscrete1, BTEdiscrete2}, some explore different solution strategies \cite{BTEKrylov}, and others leverage the parallel capabilities of equations to accelerate computations \cite{BTEacce1,BTEacce2}. 

However, most BTE methods are based on the single-material model. Although methods such as Acuros CTS \cite{AcurosCT2} incorporate high-attenuation materials such as bone, simple linear material mapping cannot accurately describe the nonlinear energy dependence of photon interaction cross-sections in different materials. In addition, BTE is a high-dimensional problem involving energy, flight angle, and spatial position \cite{cghxy}. Its accuracy and computational efficiency strongly depend on the discretization resolution of the phase space. To improve computational efficiency, large-scale downsampling operations are usually carried out. Downsampling is irreversible and will discard local material variations associated with boundaries and complex structures, which brings fine structures are averaged and affects the accuracy. In order to achieve high-precision modeling while also considering solution efficiency, this work proposed a scatter correction method based on the wavelet adaptive material-dependent BTE (WAM-BTE). The main contributions are summarized as follows.

\begin{enumerate}
	\item The conventional single-material model is extended to a multi-material model. Different attenuation coefficients, effect cross-sections, and scattering differential cross-sections are assigned to different materials. This reduces the scatter estimation bias caused by the linear approximation of a single-material model.
	\item The reversible wavelet decomposition replaces replaces irreversible downsampling operations to reduce the computational load. The low-frequency and high-frequency coefficients are simultaneously obtained. The former is used for coarse-scale scatter estimation, and the latter is used to guide corrections to adaptive local refinement.
	\item An energy selection function is formed by the high-frequency wavelet coefficients of high attenuation materials, and identify regions with larger local errors through relative thresholds. Local fine-scale correction is performed in these complex regions, which compensates for the local bias introduced by coarse low-frequency approximation. 
	\item Overall, an adaptive multi-scale scattering estimation through wavelet decomposition is established, which balances physical accuracy, low-frequency compression, and local high-frequency refinement.
\end{enumerate}

The remainder of this paper is organized as follows. Section \ref{method} describes the proposed method and presents the corresponding theoretical analysis. Section \ref{experiment} evaluates the proposed method under different experimental settings. Section \ref{discussion} discusses the results and limitations and concludes this work.

\section{Method}
\label{method}

\subsection{Material-dependent BTE scatter model}
\label{BTE model}

During CT scanning, the X-ray source continuously emits photons, and the photon distribution in the system can be regarded as approximately steady. Neglecting external forces, the photon transport process can be described by the steady-state Boltzmann transport equation according to particle conservation:
\begin{equation}
	\label{BTEall}
	\hat{\omega}\cdot\nabla \phi(\mathbf{x},E,\hat{\omega})
	=
	\mathcal{X}(\mathbf{x},E,\hat{\omega})
	+
	\mathcal{C}_{\mathrm{gain}}
	(\mathbf{x},E'\rightarrow E,\hat{\omega}'\rightarrow\hat{\omega})
	-
	\mathcal{C}_{\mathrm{loss}}
	(\mathbf{x},E,\hat{\omega}).
\end{equation}
The left-hand side of Eq. (\ref{BTEall}) describes the spatial variation of photons with energy $E$ travel along direction $\hat{\omega}$ at position $\mathbf{x}$ in phase space. Here, $\phi$ denotes the photon flux. The right-hand side contains three components: the source term, the gain term, and the loss term.

The source term $\mathcal{X}$ describes the phase-space distribution of photons emitted by the X-ray source. Its form is determined by the target material of the X-ray tube, tube voltage, and filtration parameters. It can be modeled as a directional point source with a prescribed energy distribution.

The gain term $\mathcal{C}_{\mathrm{gain}}$ describes the contribution of photons transferred from an incident state $(E',\hat{\omega}')$ to an outgoing state $(E,\hat{\omega})$. In the diagnostic X-ray energy range, Compton scattering and Rayleigh scattering are considered as the dominant scattering processes. Unlike conventional single-material equivalent BTE models, the proposed model represents the object as a combination of multiple materials:
\begin{equation}
	\label{CGain}
	\mathcal{C}_{\mathrm{gain}}
	(\mathbf{x},E'\rightarrow E,\hat{\omega}'\rightarrow\hat{\omega})
	= \int\int \sum_{m=1}^{M} \Big[
	\kappa_{\mathrm{C},m} + \kappa_{\mathrm{R},m}
	\Big] \times \phi(\mathbf{x},E',\hat{\omega}') \,d\hat{\omega}'\,dE'.
\end{equation}
Here, $m$ denotes the material index. $\kappa_{\mathrm{C},m}(\mathbf{x},E'\rightarrow E,\hat{\omega}'\rightarrow\hat{\omega})$ and $\kappa_{\mathrm{R},m}(\mathbf{x},E'\rightarrow E,\hat{\omega}'\rightarrow\hat{\omega})$ denote the Compton and
Rayleigh transfer kernels, respectively. They are determined by the local electron density and the corresponding scattering differential cross-sections:
\begin{equation}
	\label{kappaCS}
	\kappa_{\mathrm{C},m}(\mathbf{x},E'\rightarrow E,\hat{\omega}'\rightarrow\hat{\omega}) =
	\rho_{e,m}(\mathbf{x}) \cdot
	\frac{d\sigma_{\mathrm{C},m}}{d\Omega}
	(E',\hat{\omega}'\rightarrow\hat{\omega}) \cdot \delta
	\left( E-\frac{E'}{1+\vartheta} \right),
\end{equation}
\begin{equation}
	\label{kappaRS}
	\kappa_{\mathrm{R},m}(\mathbf{x},E'\rightarrow E,\hat{\omega}'\rightarrow\hat{\omega}) =
	\rho_{e,m}(\mathbf{x}) \cdot \frac{d\sigma_{\mathrm{R},m}}{d\Omega}
	(E',\hat{\omega}'\rightarrow\hat{\omega}) \cdot \delta(E-E'),
\end{equation}
\begin{equation}
	\label{vartheta}
	\vartheta = \frac{E'}{m_e c^2} \left[ 1-\cos(\hat{\omega}',\hat{\omega}) \right],
\end{equation}
where $m_e$ is the electron rest mass, $c$ is the speed of light in vacuum, $\cos(\hat{\omega}',\hat{\omega})$ denotes the cosine of the angle between the incident and outgoing directions, and $\delta(\cdot)$ is the Dirac delta function.

The scattering differential cross-sections for Compton and Rayleigh scattering in Eq. (\ref{kappaCS}) and Eq. (\ref{kappaRS}) are given by
\begin{equation}
	\label{dsigmaCS}
	\frac{d\sigma_{\mathrm{C},m}}{d\Omega} =
	\frac{r_e^2}{2} \left( \frac{1}{1+\vartheta} \right)^2
	\left[ 1+\vartheta + \frac{1}{1+\vartheta} - \sin^2(\hat{\omega}',\hat{\omega}) \right] S_m(q,Z_m),
\end{equation}
\begin{equation}
	\label{dsigmaRS}
	\frac{d\sigma_{\mathrm{R},m}}{d\Omega} = \frac{r_e^2}{2}
	\left[ 1+\cos^2(\hat{\omega}',\hat{\omega}) \right]
	\left| F_m(q,Z_m) \right|^2.
\end{equation}
Here, $r_e$ is the classical electron radius. $S_m$ and $F_m$ represent the shape correction factors corresponding to the $m$-th material, which can be queried in the classic Hubbell table \cite{Hubbell}. The variable $q$ denotes the momentum transfer determined by the photon energy and scattering angle.

\subsection{Wavelet Approximation and Adaptive Refinement}
\label{wavelet}

For a detector pixel $\mathbf{d}=(d_u,d_v)$, the scatter signal can be expressed as an integral over all scattering volume within the object domain $\Omega$:
\begin{equation}
	\label{detScatter}
	\mathcal{S}(\mathbf{d}) = I_0
	\int \int s(E) \int_{\Omega} \mathcal{G}(\mathbf{d},\mathbf{x},\beta)
	\sum_{m=1}^{M} \rho_m(\mathbf{x}) \mathcal{K}_m(\mathbf{d},\mathbf{x},E,\beta)
	\,d\mathbf{x}\,d\beta\,dE.
\end{equation}
Here, $s(E)$ denotes the normalized X-ray spectrum. $\mathcal{G}(\mathbf{d},\mathbf{x},\beta)$ is a geometric weighting term that accounts for factors such as propagation distance and solid angle $\beta$. $\mathcal{K}_m(\mathbf{d},\mathbf{x},E,\beta)$ denotes the material-dependent BTE integral kernel, whose physical components have been described in the previous subsection.

The scatter formulation can be regarded as a smooth integral transform with respect to the material density distribution $\rho_m$. A three-dimensional wavelet decomposition is therefore applied to $\rho_m$. Let $w$ denote the decomposition level and $W$ the maximum decomposition level. Let $\mathbf{n}=(n_x,n_y,n_z)$ denote the three-dimensional translation index. The material density can then be represented as
\begin{equation}
	\label{rhoWWT}
	\rho_m(\mathbf{x})
	=
	\sum_{\mathbf{n}}
	A_{m,W}(\mathbf{n})
	\Phi_{W,\mathbf{n}}(\mathbf{x})
	+
	\sum_{w=1}^{W}
	\sum_{\alpha\in\mathcal{H}}
	\sum_{\mathbf{n}}
	D_{m,w}^{\alpha}(\mathbf{n})
	\Psi_{w,\mathbf{n}}^{\alpha}(\mathbf{x}),
\end{equation}
where
\begin{equation}
	\mathcal{H} = \{LLH,LHL,LHH,HLL,HLH,HHL,HHH\}
\end{equation}
contains the seven high-frequency directional subbands in the three-dimensional wavelet decomposition. $\Phi_{W,\mathbf{n}}(\mathbf{x})$ denotes the three-dimensional scaling basis function at level $W$, and $A_{m,W}(\mathbf{n})$ is the corresponding low-frequency approximation coefficient of the $m$-th material. $\Psi_{w,\mathbf{n}}^{\alpha}(\mathbf{x})$ denotes the three-dimensional wavelet basis function at level $w$ and directional subband $\alpha$, while $D_{m,w}^{\alpha}(\mathbf{n})$ denotes the corresponding high-frequency detail coefficient.

For the Haar wavelet, after $W$ levels of three-dimensional discrete wavelet decomposition, the low-frequency coefficient can be written as
\begin{equation}
	\label{WWTlow}
	A_{m,W}(x,y,z) = \frac{1}{2^{3W/2}} \sum_{i=1}^{\ell} \sum_{j=1}^{\ell} \sum_{k=1}^{\ell} \rho_m (\ell x+i,\ell y+j,\ell z+k),
\end{equation}
where $\ell=2^W$ is the coarse-scale factor at level $W$. 

Substituting the wavelet decomposition (Eq. (\ref{rhoWWT})) into the scatter model (Eq. (\ref{detScatter})), the scatter signal can be decomposed into low-frequency and high-frequency components:
\begin{equation}
	\label{sl+sh}
	\mathcal{S}(\mathbf{d}) = 
	\mathcal{S}_{L}(\mathbf{d}) + \mathcal{S}_{H}(\mathbf{d}),
\end{equation}
\begin{equation}
	\label{sl}
	\begin{aligned}
		\mathcal{S}_{L}(\mathbf{d})
		&=
		I_0
		\int\int
		s(E)
		\int_{\Omega}
		\mathcal{G}(\mathbf{d},\mathbf{x},\beta)
		\sum_{m=1}^{M}
		\sum_{\mathbf{n}}
		A_{m,W}(\mathbf{n})
		\Phi_{W,\mathbf{n}}(\mathbf{x}) \\
		&\qquad\qquad\times
		\mathcal{K}_m
		(\mathbf{d},\mathbf{x},E,\beta)
		\,d\mathbf{x}\,d\beta\,dE,
	\end{aligned}
\end{equation}
\begin{equation}
	\label{sh}
	\begin{aligned}
		\mathcal{S}_{H}(\mathbf{d})
		&=
		I_0
		\int\int
		s(E)
		\int_{\Omega}
		\mathcal{G}(\mathbf{d},\mathbf{x},\beta)
		\sum_{m=1}^{M}
		\sum_{w=1}^{W}
		\sum_{\alpha\in\mathcal{H}}
		\sum_{\mathbf{n}}
		D_{m,w}^{\alpha}(\mathbf{n}) \\
		&\qquad\qquad\times
		\Psi_{w,\mathbf{n}}^{\alpha}(\mathbf{x})
		\mathcal{K}_m
		(\mathbf{d},\mathbf{x},E,\beta)
		\,d\mathbf{x}\,d\beta\,dE.
	\end{aligned}
\end{equation}

When the path length through high attenuation materials is short or their spatial structures are smooth, the dominant scatter contribution can be approximated by the low-frequency component $ \mathcal{S}(\mathbf{d})\approx \mathcal{S}_{L}(\mathbf{d})$. 

However, when the path length through high attenuation materials is large or when abrupt and complex structures are present, the coarse low-frequency approximation may introduce considerable local scatter errors. To identify these regions, the high-frequency wavelet coefficients of high attenuation materials are used to define a local structural complexity measure. For an arbitrary coarse block $V$, the material-weighted high-frequency energy is defined as
\begin{equation}
	\label{energyFunc}
	E(V) = \sum_{m=1}^{M} \gamma_m \sum_{\alpha\in\mathcal{H}}
	\sum_{\mathbf{n}\in V} \left| D_{m,w}^{\alpha}(\mathbf{n}) \Psi_{w,\mathbf{n}}^{\alpha}(\mathbf{x}) \right|^2,
\end{equation}
where $\gamma_m$ denotes the material weight used to emphasize materials that are more sensitive to scatter estimation errors.

\subsection{Compared with downsampling strategy}
\label{compareDS}

Considering (Eq. (\ref{detScatter})), the scatter formulation can be regarded as a smooth integral transform with respect to the material density distribution $\rho_m$. Obviously, it is a high-dimensional computational problem about x-ray energy, detectors, scattering distribution and the density of the scanned object.

To reduce the computational load, a common strategy is to approximate the full-resolution material density $\rho_{m,L}$ using the downsampling block-averaged density $\bar{\rho}_m$ on a coarse voxel
grid. Let the coarse voxel size be $\ell\times\ell\times\ell$. The three-dimensional block-averaged density is written as
\begin{equation}
	\label{barrho}
	\bar{\rho}_m(x,y,z) = \frac{1}{\ell^3}
	\sum_{i=1}^{\ell} \sum_{j=1}^{\ell} \sum_{k=1}^{\ell} \rho_m (\ell x+i,\ell y+j,\ell z+k),
\end{equation}
where $(x,y,z)$ denotes the coarse-grid voxel index. Obviously, at the same scale $\ell = 2^W$, the downsampling block-averaged approximation and the low-frequency wavelet approximation differ by only one constant, i.e.,
\begin{equation}
	\bar{\rho}_m = \ell^{-3/2} A_{m,W}
\end{equation}
and the constant is in the three-dimensional scaling basis function $\Phi_{W,\mathbf{n}}(\mathbf{x})$.

\section{Experiment}
\label{experiment}

All experiments are performed on the same workstation equipped with a 3.70 GHz Intel Ultra 7 270K Plus 24-core CPU and an NVIDIA GeForce RTX 5080 GPU. The same computational platform is used for all methods to ensure a fair comparison of computational efficiency.

\begin{figure}
	\centering
	\includegraphics[width=0.95\textwidth]{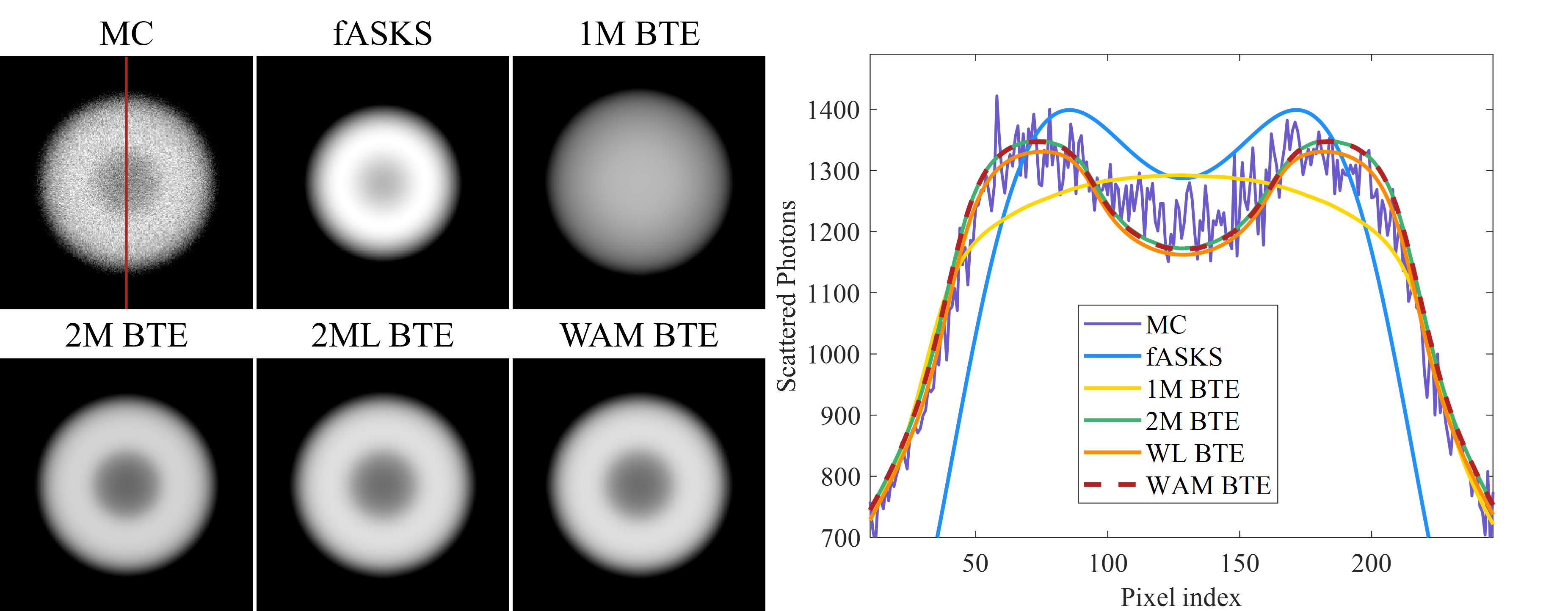}
	\caption{The estimated scatter distribution and the corresponding profiles of the sphere phantom. Display window: [1000, 1400] photons.}
	\label{fig-sphere-scatter}
\end{figure}

Fig. \ref{fig-sphere-scatter} shows the scatter distributions estimated by different methods. The MC result exhibits a clear radially nonuniform distribution, reflecting the different attenuation and scattering properties of the inner and outer materials. faSKS captures the overall low-frequency trend but shows a systematic deviation from the MC reference. The 1M-BTE model also exhibits noticeable errors because a single equivalent material cannot accurately describe the scattering properties of different material regions. Compared with the MC, the result of 2M-BTE is substantially improved, demonstrating the importance of material-dependent modeling. Although 2ML-BTE uses a coarse-scale material representation, its overall scatter distribution remains close to that of the full-resolution 2M-BTE. This result confirms that the dominant scatter information can be preserved by the low-frequency representation. WAM-BTE further corrects local deviations using high-frequency-guided refinement and provides a scatter profile closer to MC and 2M-BTE.

\begin{figure}
	\centering
	\includegraphics[width=0.95\textwidth]{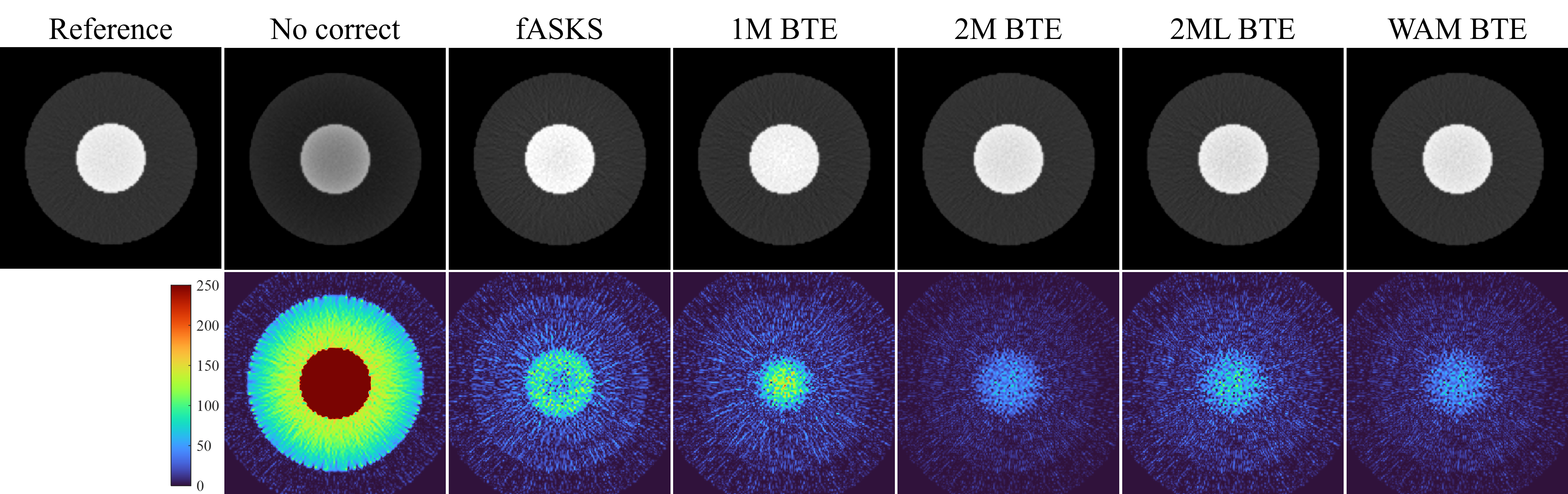}
	\caption{The corrected reconstruction results and absolute error maps of the sphere phantom. The display window of the first row is set to [0.01, 0.04]. To better observe the correction effect, the second row results are transformed to CT values, with the window width [0, 250] Hu.}
	\label{fig-sphere-rec}
\end{figure}

Figs. \ref{fig-sphere-rec} shows the corresponding reconstruction results and absolute error maps. Without scatter correction, a pronounced low-frequency intensity bias appears over the entire phantom. faSKS and 1M-BTE reduce this bias but retain visible residual errors. The errors are substantially reduced by the material-dependent BTE models. WAM-BTE provides a reconstruction close to the reference while maintaining accurate values near the material boundaries.

\subsection{Simulated medical phantoms}
\label{results-med}

\begin{figure}
	\centering
	\includegraphics[width=0.95\textwidth]{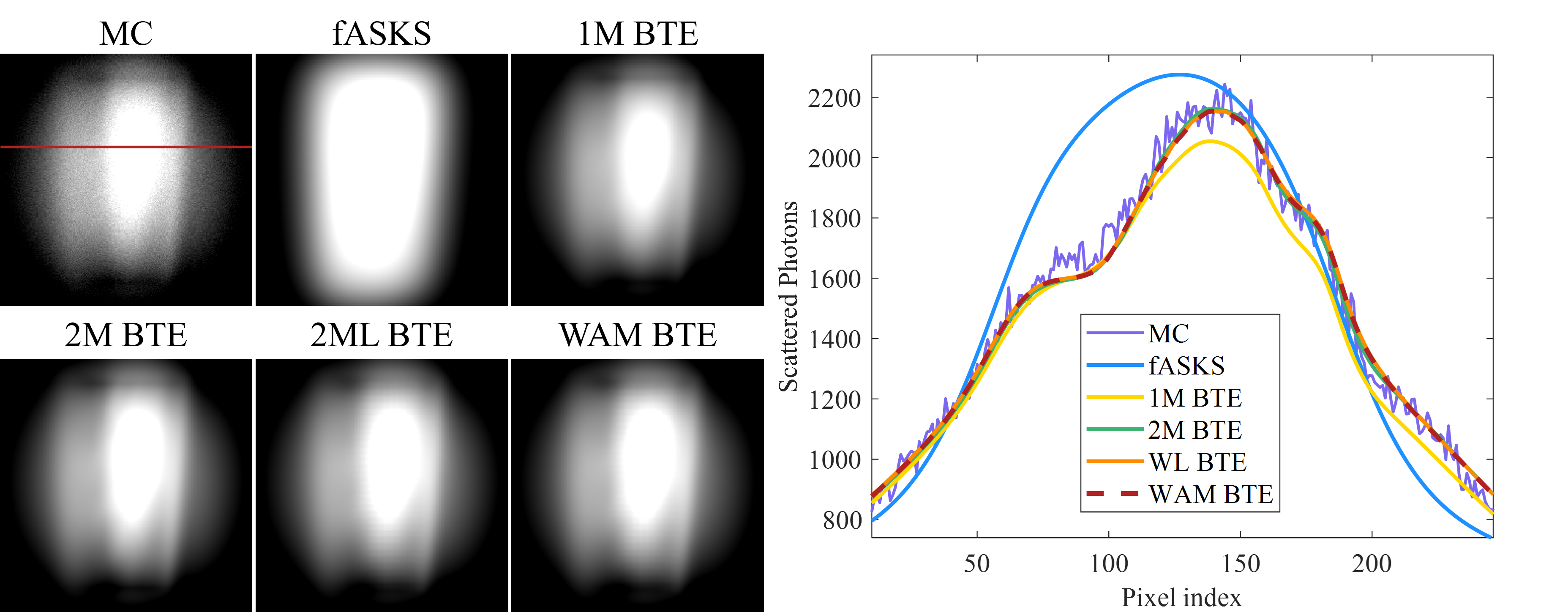}
	\caption{The estimated scatter distribution and the corresponding profiles of the thorax phantom. Display window: [900, 1900] photons.}
	\label{fig-thorax-scatter}
\end{figure}

\begin{figure}
	\centering
	\includegraphics[width=0.95\textwidth]{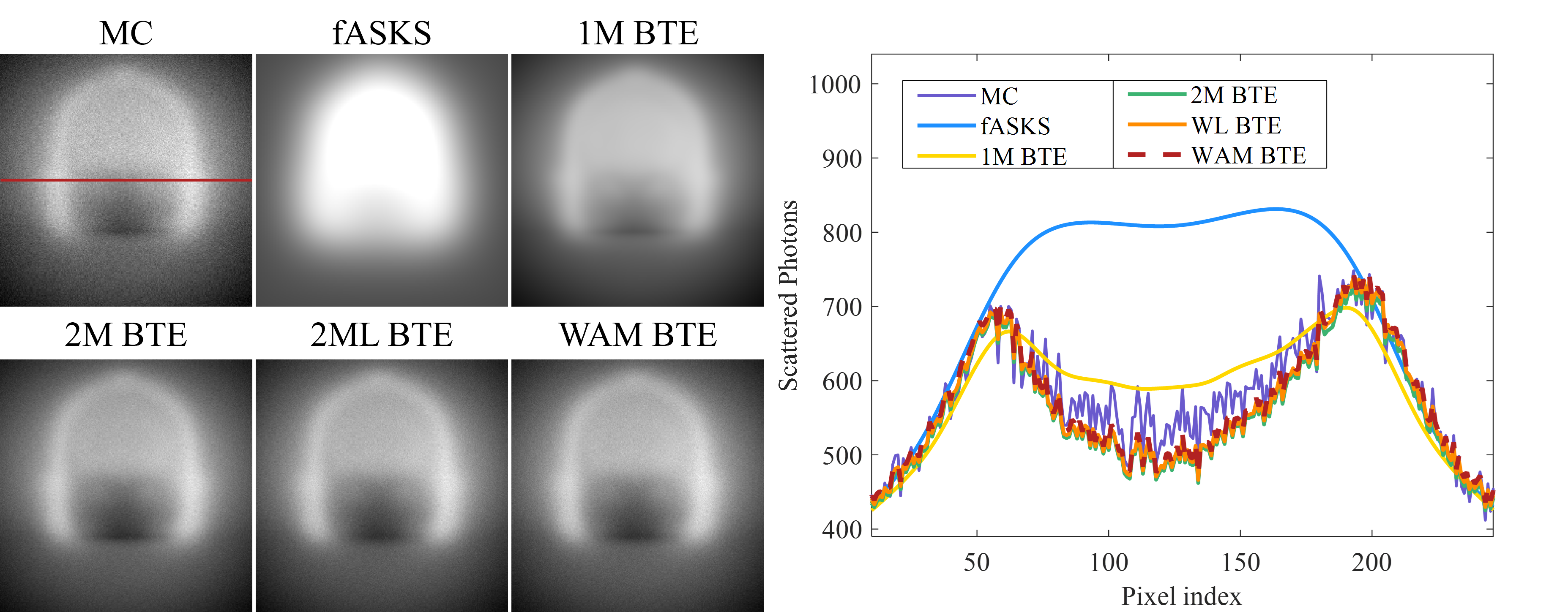}
	\caption{The estimated scatter distribution and the corresponding profiles of the head phantom. Display window: [200, 800] photons.}
	\label{fig-head-scatter}
\end{figure}

Figs. \ref{fig-thorax-scatter} and \ref{fig-head-scatter} show the scatter estimation results for the thorax and head phantoms, respectively. Compared with the sphere phantom, these medical phantoms contain more complex anatomical structures, irregular material boundaries, and heterogeneous distributions of low attenuation and high attenuation materials.

For the thorax phantom, MC can clearly show the influence of different tissues, like lungs, soft tissues, and bone, on the scattering distribution. faSKS produces an excessively smooth distribution and overestimates the scatter level in the central region. Since most of the components in the thorax phantom are low attenuation materials, 1M-BTE model recovers part of the spatial variation but still exhibits noticeable deviations around high attenuation structures. Explicit modeling of different materials in 2M-BTE substantially improves the agreement with the MC reference. 2ML-BTE preserves the main spatial characteristics after coarse-scale compression, whereas WAM-BTE further reduces local deviations in structurally complex regions.

A similar tendency is observed for the head phantom. The presence of dense and complex bone structures results in stronger local variations in the scatter distribution. faSKS fails to reproduce the central depression and local variations observed in the MC reference, while 1M-BTE still exhibits material-related model bias. In contrast, the material-dependent BTE models better reproduce the scatter distribution around the bone structures. WAM-BTE maintains good agreement with the MC reference over most detector positions.

\begin{figure}
	\centering
	\includegraphics[width=0.95\textwidth]{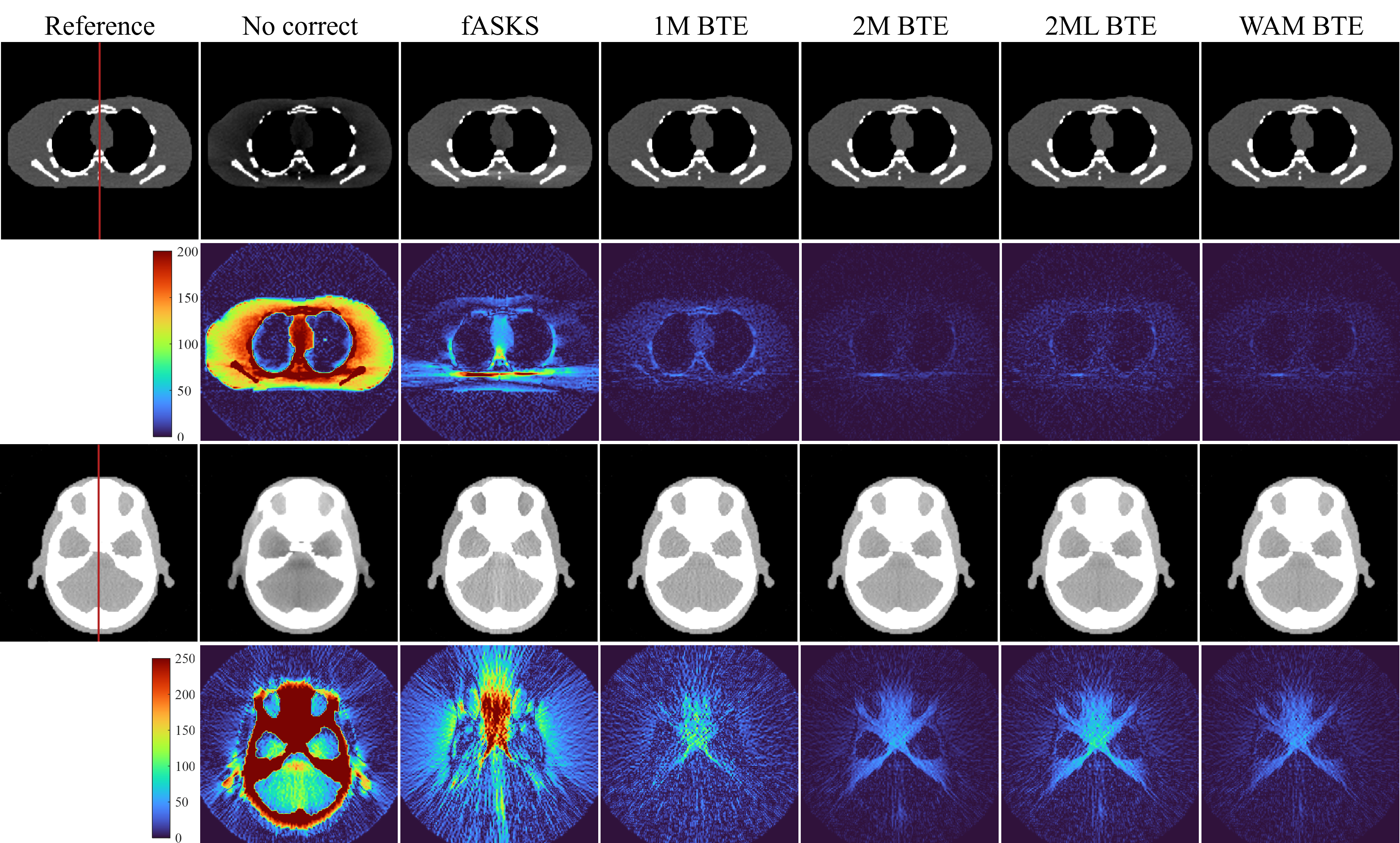}
	\caption{The corrected reconstruction results and absolute error maps of the thorax and head phantom. The display windows of the first and third rows are respectively set to [0.01, 0.03] and [0.00, 0.02]. To better observe the correction effect, the absolute error maps are transformed to CT values, with the window width [0, 200] and [0, 250] Hu respectively.}
	\label{fig-thorax-rec}
\end{figure}

\begin{figure}
	\centering
	\includegraphics[width=0.45\textwidth]{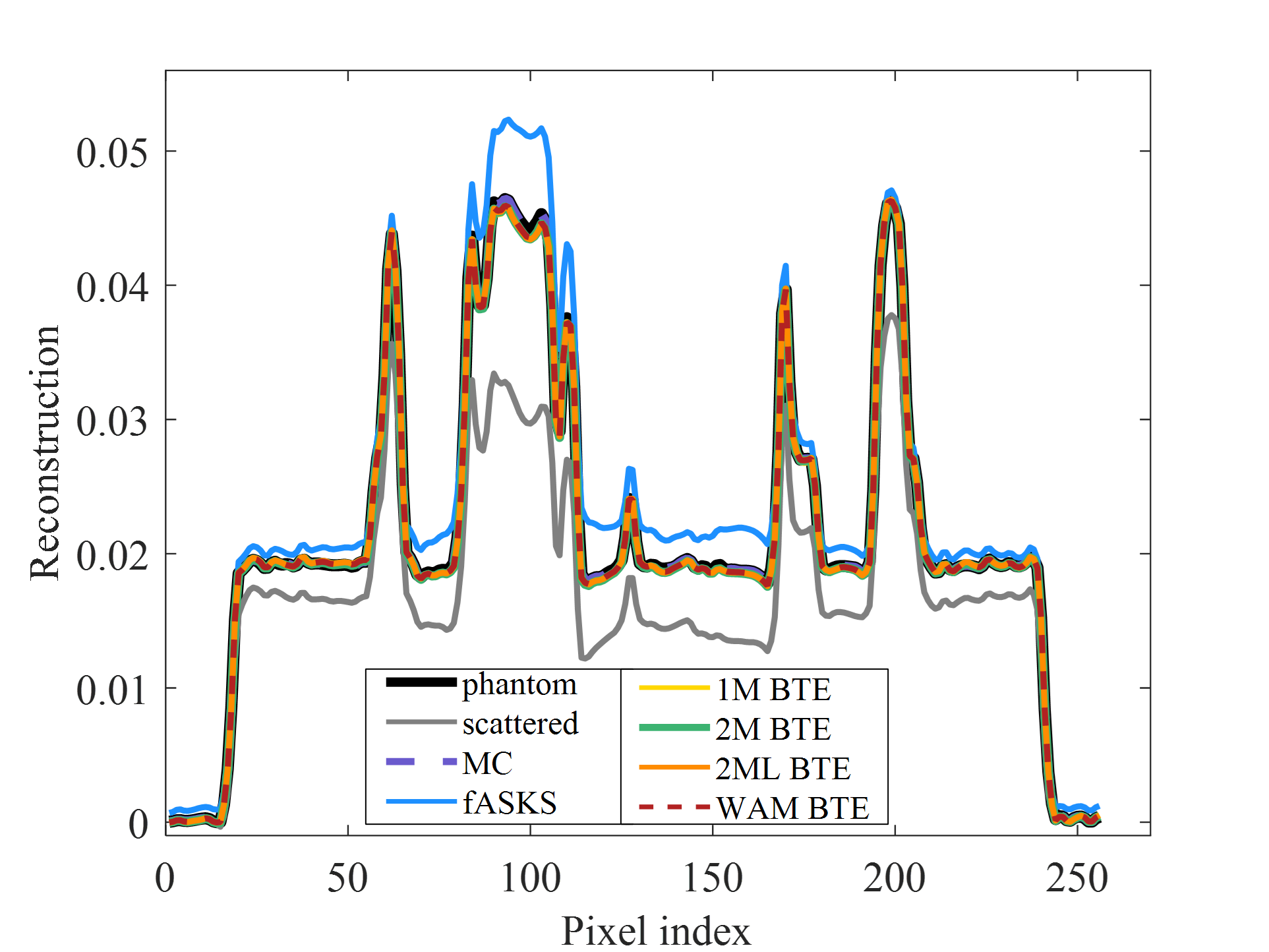}
	\includegraphics[width=0.45\textwidth]{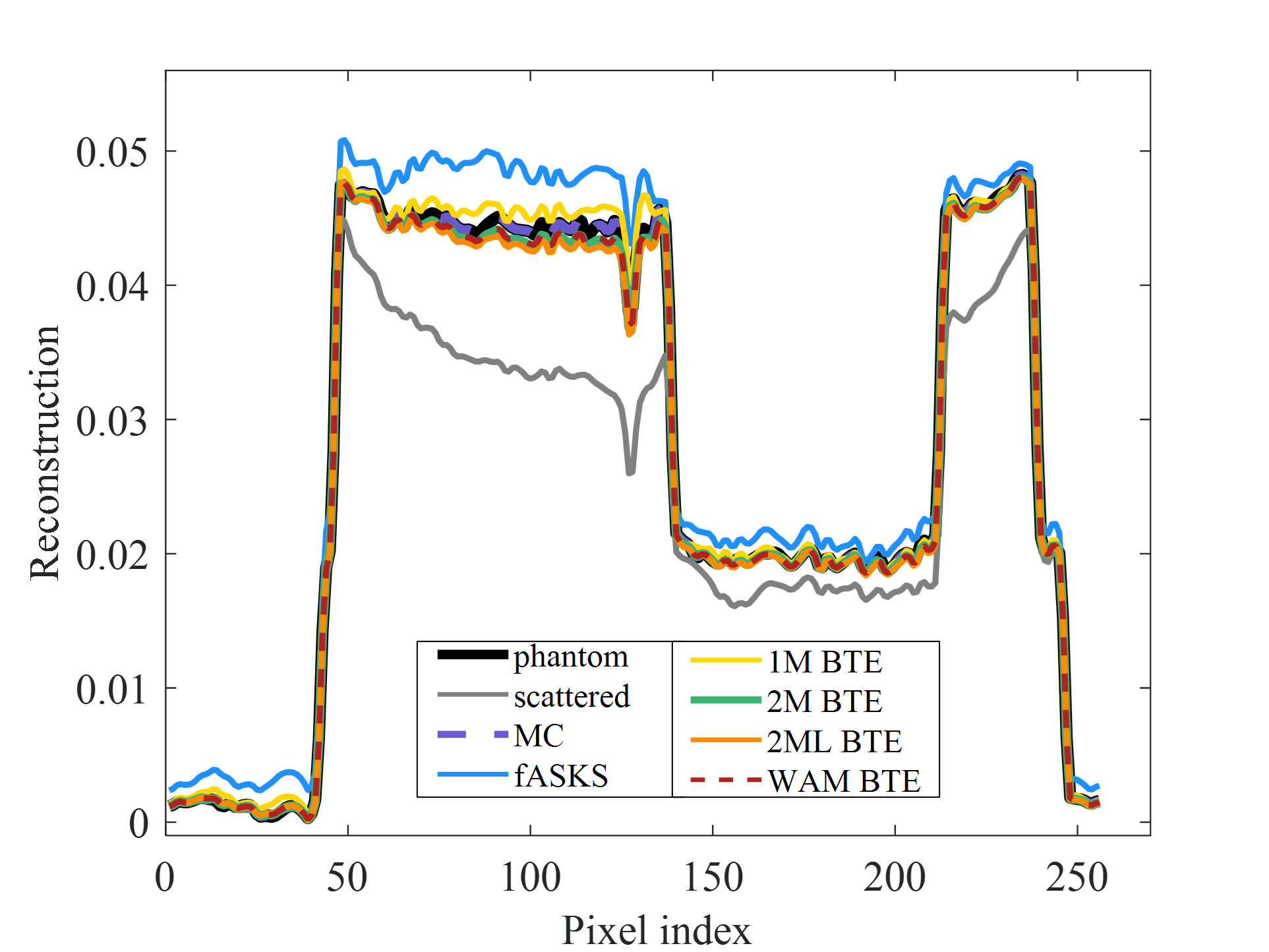}
	\caption{The profiles of the thorax and head phantoms.}
	\label{fig-thorax-rec-plot}
\end{figure}

Figs. \ref{fig-thorax-rec} shows the reconstructed images and absolute error maps, and Fig. \ref{fig-thorax-rec-plot} is the corresponding profiles. faSKS and 1M-BTE reduce the overall scatter-induced bias but leave evident structure-dependent residual errors. These errors are significantly reduced by 2M-BTE, particularly near interfaces between bone and soft tissue. Although 2ML-BTE uses a coarse spatial representation, it retains most of the correction performance of the full-resolution model. WAM-BTE further compensates for local approximation errors while preserving the computational advantage of the coarse-scale model.

\subsection{Computational efficiency}
\label{results-time}

The computational efficiency of different scatter estimation methods is further evaluated to verify the theoretical analysis in Section \ref{acc}. Table \ref{tab-time} reports the average scatter estimation time for per projection view. All methods were evaluated on the same computational platform.

\begin{table}[!t]
	\centering
	\caption{Average calculation time of scatter estimation for per projection view.}
	\label{tab-time}
	\renewcommand{\arraystretch}{1.15}
	\setlength{\tabcolsep}{4.5pt}
	\begin{tabular}{lcc}
		\hline
		\textbf{Method} &
		\textbf{Simulation (s)} &
		\textbf{Real data (s)} \\
		\hline
		MC      & 15.323 & 33.010 \\
		faSKS   & 0.022 & 0.018 \\
		1M BTE  & 5.501 & 12.512 \\
		2M BTE  & 8.004 & 20.655 \\
		2ML BTE & 0.027 & 1.584 \\
		WAM BTE & 2.125 & 8.034 \\
		\hline
	\end{tabular}
\end{table}

Among all methods, MC simulation requires the longest computation time, with 15.323 s and 33.010 s per projection view for the simulation and real-data experiments, respectively. This is expected because MC methods require the transport histories of a large number of X-ray photons to be sampled statistically to obtain a sufficiently stable scatter distribution. In contrast, faSKS provides the shortest computation time because it evaluates a simplified analytical scatter model without explicitly solving photon transport.

As discussed in Section \ref{acc}, introducing multiple materials does not change the asymptotic order of the BTE solver, but increases the constant computational cost. This trend is also observed in Table \ref{tab-time}. For the simulation experiment, the computation time increases from 5.501 s for 1M BTE to 8.004 s for 2M BTE. For the real-data experiment, the corresponding time increases from 12.512 s to 20.655 s. Although the two-material model introduces additional material-dependent calculations, the runtime does not increase by a factor of two. This is because the computations for different materials can be parallelized, while several geometric quantities and transport-related operations are shared among materials.

A two-level three-dimensional wavelet decomposition, i.e., $W=2$, is used for both experiments. According to Eq. (\ref{compuLow}), the low-frequency representation reduces the number of spatial samples to $1/2^{3W}=1/64$ of the original resolution before adaptive refinement. As shown in Table \ref{tab-time}, the measured computation times are consistent with the theoretical complexity trend. For the simulation experiment, the full-resolution 2M BTE requires 8.004 s per projection view, whereas 2ML BTE requires only 0.027 s. This substantial reduction confirms that most of the computational burden of BTE scatter estimation can be removed by performing the dominant scatter calculation on the low-frequency volume. Although the actual speedup does not strictly follow the theoretical $2^{3W}$ scaling because of parallel computation, shared geometric operations, memory access, and other implementation overheads, the reduction in runtime clearly demonstrates the computational benefit predicted by Eq. (\ref{compuLow}).

The adaptive refinement in WAM-BTE increases the computation time to 2.125 s in the simulation experiment. However, it remains substantially faster than the full-resolution 2M BTE, reducing the runtime from 8.004 s to 2.125 s, corresponding to a reduction of approximately 73.5\% and a speedup of about 3.77 times. This behavior is consistent with Eq. (\ref{compuAd}), i.e., the proposed method restores fine-scale calculations only in wavelet-activated regions rather than over the entire volume. Therefore, local scatter estimation accuracy can be improved with a considerably smaller computational cost than that required by full-resolution refinement.

A similar trend is observed for the larger real-data experiment. The computation time of 2M BTE increases to 20.655 s per view because of the substantially larger reconstruction volume, whereas 2ML BTE requires only 1.584 s. WAM-BTE requires 8.034 s, which is still approximately 61.1\% lower than that of 2M BTE and corresponds to a speedup of about 2.57 times.

Notably, 2ML BTE achieves a computation time of only 27 ms per projection view in the simulation experiment while retaining the dominant low-frequency scatter distribution. Therefore, when extremely high local correction accuracy is not required, the coarse-scale BTE model provides an efficient option for practical scatter estimation. The goal of WAM-BTE is to strike a balance between efficiency and accuracy. It introduces additional computation only in structurally complex regions identified by the high-frequency wavelet components. The results demonstrate that this local refinement improves the local modeling accuracy at a much lower computational cost than restoring the full-resolution BTE calculation over the entire volume.

\section*{Reference}
\bibliographystyle{unsrt}
\bibliography{references}

\end{document}